\documentclass[12pt]{article}
\usepackage{times}
\usepackage{epsfig}

\def\beq{\begin{equation}}
\def\eeq{\end{equation}}
\def\bea{\begin{eqnarray}}
\def\eea{\end{eqnarray}}
\def\d{\displaystyle}

\def\nn{\nonumber}
\def\Eq#1{Eq.~(\ref{#1})}

\begin{document}
\begin{titlepage}

\renewcommand{\thefootnote}{\fnsymbol{footnote}}
\begin{flushright}
    IFIC/07-51   \\ TTP07-25 \\ arXiv:0709.1652 [hep-ph]
\end{flushright}
\par \vspace{10mm}
  
\begin{center}

{\Large \bf Top Quarks, Axigluons and Charge Asymmetries \\
 at Hadron Colliders}

\vspace{8mm}

{\bf Oscar Antu\~nano~$^{(a)}$\footnote{E-mail: oscant@ific.uv.es}},
{\bf Johann H. K\"uhn~$^{(b)}$\footnote{E-mail: johann.kuehn@uni-karlsruhe.de}},
{\bf Germ\'an Rodrigo~$^{(a)}$\footnote{E-mail: german.rodrigo@ific.uv.es}}

\vspace{5mm}
${}^{(a)}$ Instituto de F\'{\i}sica Corpuscular, 
CSIC-Universitat de Val\`encia, \\
Apartado de Correos 22085, 
E-46071 Valencia, Spain. \\
\vspace{5mm} 
${}^{(b)}$ Institut f\"ur Theoretische Teilchenphysik,
Universit\"at Karlsruhe, D-76128 Karlsruhe, Germany. \\

\vspace{5mm}
\end{center}

\par \vspace{2mm}
\begin{center} {\large \bf Abstract} \end{center}
\begin{quote}
\pretolerance 10000
Axigluons are colored heavy neutral gauge boson that couple to 
quarks through an axial vector current and the same strong 
coupling as gluons. The most important 
model-independent manifestation of axigluons is the generation 
of a forward--backward asymmetry in top-antitop quark production 
at $p\bar{p}$ collisions which originates from the charge asymmetry. 
We update our previous analysis for the inclusive QCD induced 
forward--backward asymmetry and define a new observable which 
is more sensitive to the effect than the forward--backward asymmetry. 
Furthermore, we find a lower limit of $1.2$~TeV at 90\% C.L.
on the axigluon mass from recent measurements of the asymmetry 
at Tevatron. Also at LHC, the charge asymmetry is sizable in suitably
selected samples. We evaluate this asymmetry 
in the central region for different selection cuts and show that, 
like at Tevatron, the charge asymmetry can probe larger values 
of the axigluon mass than the dijet mass distribution. 
\end{quote}

\vspace*{\fill}
\begin{flushleft}
     IFIC/07-51 \\ 
     TTP07-25 \\
     September 11, 2007
\end{flushleft}
\end{titlepage}

\setcounter{footnote}{1}
\renewcommand{\thefootnote}{\fnsymbol{footnote}}


\section{Introduction}

The Large Hadron Collider (LHC) will enter into operation very soon, 
allowing to explore the existence of new physics at the TeV energy scale with 
unprecedented huge statistics~\cite{Gianotti:2005fm}.
Since the top quark is the heaviest known 
elementary particle it plays a fundamental role in many extensions of the 
Standard Model (SM), and its production and decay channels are 
promising probes of new physics. The total cross section of top-antitop 
quark production at LHC is about $100$ times larger than at Tevatron. 
This will lead to the production of millions of $t\bar{t}$ pairs per year 
even at the initial low luminosity of ${\cal L}=10^{33}$cm$^{-2}$s$^{-1}$ 
(equivalent to $10$~fb$^{-1}$/year integrated luminosity). 

Some properties of the top quark can be studied at Tevatron 
through the forward--backward asymmetry which originates from 
the charge asymmetry~\cite{mynlo,Halzen:1987xd}. 
The Born processes relevant for top quark 
production, $q\bar{q} \to t\bar{t}$ and $gg \to t\bar{t}$, do not 
discriminate between final quark and antiquark, thus predicting 
identical differential distributions also for the hadronic 
production process. At order $\alpha_s^3$ however a charge 
asymmetry is generated and the differential distributions
of top quarks and antiquarks are no longer equal. 
A similar effect leads also to a strange-antistrange 
quark asymmetry, $s(x)\neq \bar{s}(x)$, through next-to-next-to-leading 
(NNLO) evolution of parton densities~\cite{Catani:2004nc}.
The inclusive charge asymmetry has its origin in two different reactions:
radiative corrections to quark-antiquark annihilation (Fig.~\ref{fig:qqbar})
and interference between different amplitudes contributing 
to gluon-quark scattering 
$qg \to t \bar{t}q$ and $\bar{q}g \to t \bar{t}\bar{q}$.
Gluon-gluon fusion remains obviously symmetric. The integrated 
forward--backward asymmetry has been predicted 
to be about $+5\%$ at Tevatron~\cite{mynlo}; i.e. top quarks 
are preferentially emitted in the direction of the incoming protons.  
This prediction suffers, however, from a sizable uncertainty because, 
although arising from a one-loop calculation and the corresponding real 
emission terms, it is still a leading order (LO) result. 
At LHC the total forward--backward asymmetry 
vanishes trivially because the proton-proton initial state is symmetric. 
A charge asymmetry is, however, still visible in
suitably defined distributions~\cite{mynlo}.  

In $t\bar{t}$ plus jet production, which represents an important 
background process for Higgs boson searches in electroweak vector boson 
fusion and $t\bar{t}H$ production, the asymmetry, of order $\alpha_s^3$, 
is a tree-level effect~\cite{Halzen:1987xd} with   
sign opposite to the one given for inclusive production~\cite{mynlo}.  
The next-to-leading (NLO) order QCD corrections to 
the $t\bar{t}$+jet exclusive channel
have become available very recently~\cite{ttjetnlo}, providing 
a NLO prediction for the charge asymmetry. At Tevatron the exclusive 
forward--backward asymmetry of top quarks is drastically 
reduced at NLO, from about $-7\%$ at LO to $-1.5\pm 1.5\%$,
where the large uncertainty is due to the residual scale 
dependence. A priori one can not extrapolate this unexpected 
result to the totally inclusive asymmetry.
Already at LO the size of the asymmetry in the $t\bar{t}$+jet event 
sample depends on the jet resolution parameters, and on the 
softness of the radiated gluon. 
A smaller minimal gluon energy generates a larger 
negative asymmetry, and vice versa, and the size of NLO corrections 
might be different for different jet setups. Furthermore, because
the one-loop corrections to the inclusive asymmetry are positive 
and larger than the negative contribution of real gluon 
bremsstrahlung a smaller prediction for the asymmetry in
the $t\bar{t}$+jet sample might give rise to a larger inclusive 
asymmetry. Unfortunately, the only way to obtain a more accurate 
prediction for the inclusive asymmetry is to evaluate the higher order 
corrections corresponding to a difficult two-loop calculation.

The forward--backward asymmetry of top quarks has already attracted much  
experimental interest at Tevatron. A detailed study of the feasibility of 
the measurement in the dilepton and lepton+jet channels has been presented 
in Ref.~\cite{Bowen:2005ap}. Experimental measurements have already been 
performed in Refs.~\cite{Weinelt:2006mh,Hirschbuehl:2005bj,Schwarz:2006ud}.
In Ref.~\cite{Schwarz:2006ud}, based on $695$~pb$^{-1}$ integrated
luminosity, the forward--backward asymmetry in the lepton plus jets 
channels was found to be 
\beq
A_\mathrm{FB}= 0.20 \pm 0.11~(\rm{stat}) \pm 0.047~(\rm{sys})~.
\eeq
A very similar result was obtained for the inclusive asymmetry 
in Refs.~\cite{Weinelt:2006mh,Hirschbuehl:2005bj}
based on $955$~pb$^{-1}$ integrated luminosity, 
and again in the lepton plus jets channels:
\beq
A(\Delta y \cdot Q_l)= 0.23 \pm 0.12~(\rm{stat}) 
\pm~^{0.056}_{0.057}~(\rm{sys})~,
\eeq
where the charge asymmetry is defined by the difference in the 
number of events with positive and negative $\Delta y\cdot Q_l$, 
the rapidity difference of the semileptonically and hadronically 
decaying top quark times the charge of the charged lepton. 
They also have measured the exclusive asymmetry of the four- and 
five-jet samples:
\bea
A^{\rm{4j}}(\Delta y \cdot Q_l) &=& 0.11 \pm 0.14~(\rm{stat}) 
\pm~^{0.036}_{0.034}~(\rm{sys})~. \nn \\
A^{\rm{5j}}(\Delta y \cdot Q_l) &=& 0.37 \pm 0.30~(\rm{stat}) 
\pm~^{0.075}_{0.066}~(\rm{sys})~.
\eea
The asymmetry in the five-jet sample although expected to be 
negative is statistically limited, and the comparison 
with the NLO prediction~\cite{ttjetnlo} would require the use of 
the same jet definition. 
The inclusive asymmetry in both experimental analyses, although
compatible with the theoretical prediction, is still 
statistically dominated. The statistical error, however, 
is expected to be reduced to $0.04$ with 
$8$~fb$^{-1}$~\cite{Schwarz:2006ud}, 
which is comparable with the systematic error.

With these results at hand one might try to test new 
physics beyond the SM. The production of $t\bar{t}$ through a colored 
heavy neutral gauge boson will manifest itself through a bump in 
the invariant mass distribution of the top-antitop quark 
pair~\cite{chiralcolor,Bagger:1987fz,Choudhury:2007ux}, and in 
some models might give rise to a sizable forward--backward 
asymmetry~\cite{Sehgal:1987wi}. 
Models which extend the standard color gauge group 
to $SU(3)_L \times SU(3)_R$ at high energies, 
the so called chiral color theories~\cite{chiralcolor}, 
predict the existence of a massive, color-octet 
gauge boson, the axigluon, which couples to quarks 
with an axial vector structure and the same strong 
interaction coupling strength as QCD.
Although there are many different implementations of 
chiral color theories with new particles in varying representations
of the gauge groups, the most important model-independent 
prediction of these models is the existence of the axigluon.
Similar states are also predicted in technicolor 
models~\cite{technicolor}. 
The main signature is the appearance of a 
charge asymmetry of order $\alpha_s^2$.
Because the coupling of the axigluon to quarks is an axial 
vector coupling the charge asymmetry that can be generated
is maximal. 

Axigluon masses below $1$~TeV are already ruled out. 
The Tevatron searches for new resonances decaying to 
dijets~\cite{Giordani:2003ib,Abe:1997hm} exclude the 
mass range $200 < m_{A} < 1130$ GeV at $95\%$ C.L. 
Lighter axigluons, with masses lower than $50$~GeV, 
were excluded studying $Z$ decays, where the axigluon is 
produced by the bremsstrahlung of the quarks~\cite{Rizzo:1987rm}, 
and also studying the decay of Upsilon~\cite{Cuypers:1988cb} 
and the Z decay to a gluon and an axigluon \cite{Carlson:1987ap}. 
The intermediate mass range window has also been 
excluded~\cite{Doncheski:1998ny}.

In this paper we shall update our prediction for the inclusive 
forward--backward asymmetry at Tevatron, and propose a new 
differential distribution that enhances the asymmetry by almost 
a factor $1.5$. We then obtain a new constraint  
on the axigluon mass from the actual measurement of the inclusive 
asymmetry at Tevatron. Finally, we show that at LHC the 
study of the differential charge asymmetry is sensitive to larger 
axigluon masses than the top-antitop invariant mass distribution.

\section{The QCD induced charge asymmetry}

\begin{figure}[th]
\begin{center}
\includegraphics[width=9cm]{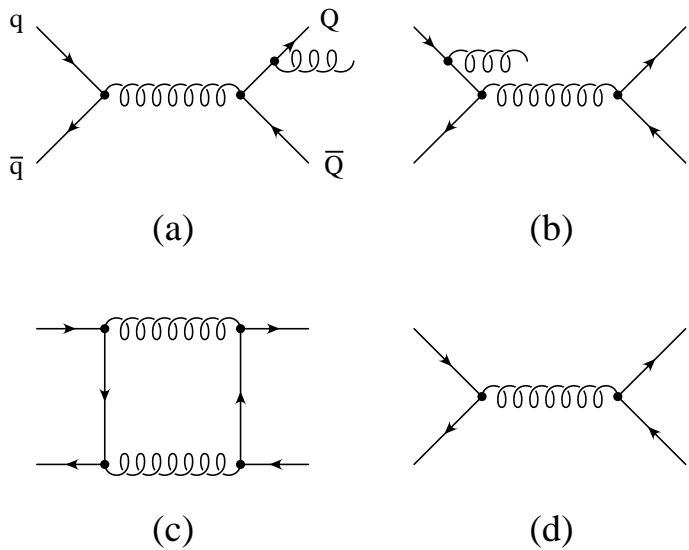}
\caption{Origin of the QCD charge asymmetry in hadroproduction 
of heavy quarks: interference of final-state 
(a) with initial-state (b) gluon bremsstrahlung,
plus interference of the double virtual gluon exchange (c) 
with the Born diagram (d). Only representative diagrams are shown.}
\label{fig:qqbar}
\end{center}
\end{figure}

The QCD induced charge asymmetry in the reaction $q\bar{q} \to t\bar{t} (g)$
is generated by the interference of final-state with initial-state 
gluon radiation  [Fig.~\ref{fig:qqbar}, (a)$\times$(b)] and 
by the interference of virtual box diagrams with 
the Born process [Fig.~\ref{fig:qqbar}, (c)$\times$(d)].
The asymmetric contribution of the virtual corrections 
exhibit soft singularities that are canceled by the 
real contribution, but do not exhibit collinear light quark mass 
singularities which would have to be absorbed by the lowest 
order process which however is symmetric. 
Ultraviolet divergences are absent for the same reason. 
The virtual plus soft radiation on one hand and the real 
hard radiation on the other contribute with opposite signs, 
with the former always larger that the latter such that 
the inclusive asymmetry becomes positive. Top quarks are 
thus preferentially emitted in the direction of the 
incoming quark at the partonic level, which translates to 
a preference in the direction of the incoming proton 
in $p\bar{p}$ collisions.  
Flavour excitation $gq(\bar{q}) \to t\bar{t}X$ generates
already at tree-level a forward--backward asymmetry which at Tevatron 
is also positive although one order of magnitude smaller than 
the asymmetry from $q\bar{q}$ annihilation.

\begin{figure}[th]
\begin{center}
\includegraphics[width=8cm]{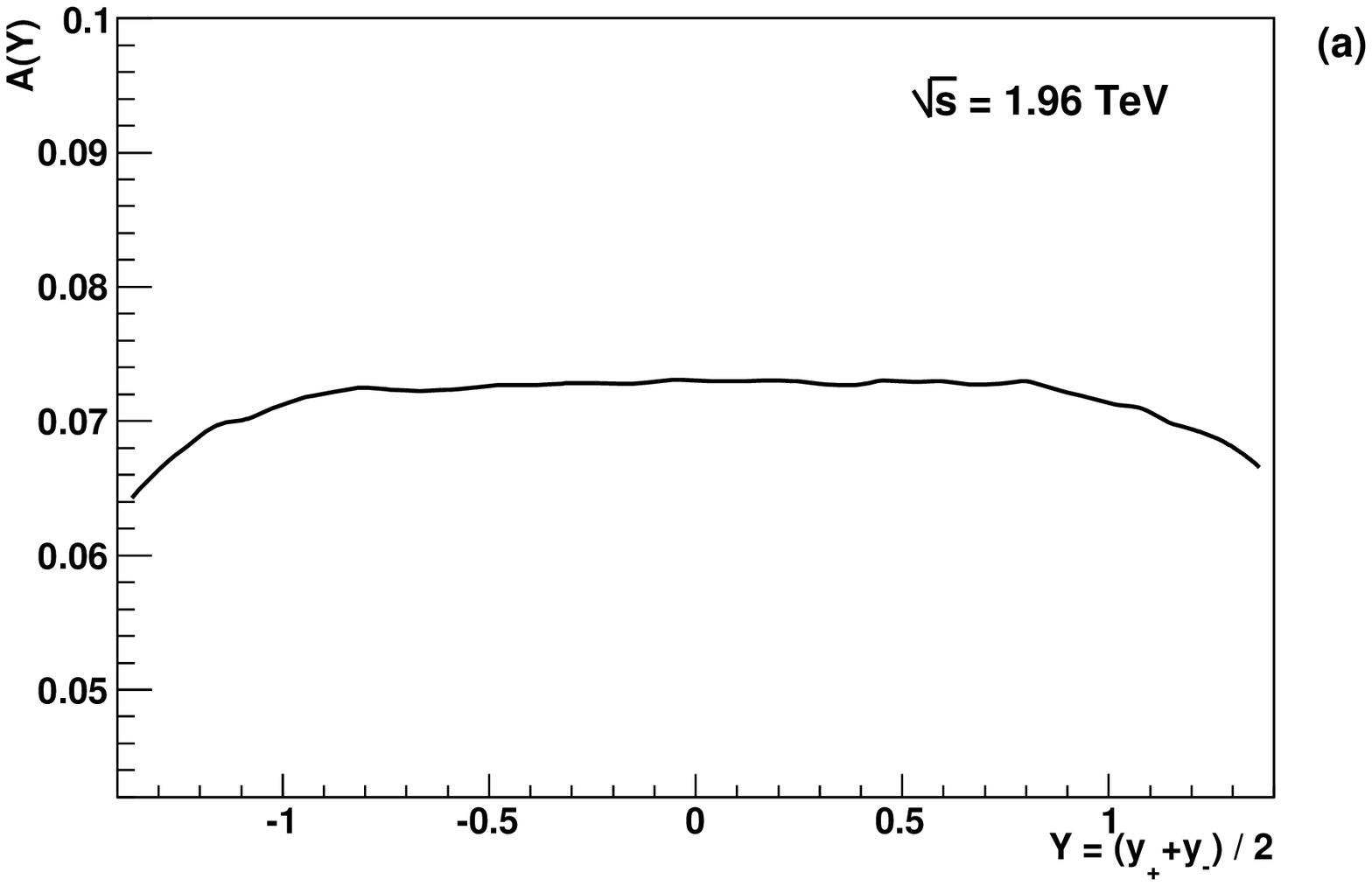} 
\includegraphics[width=8cm]{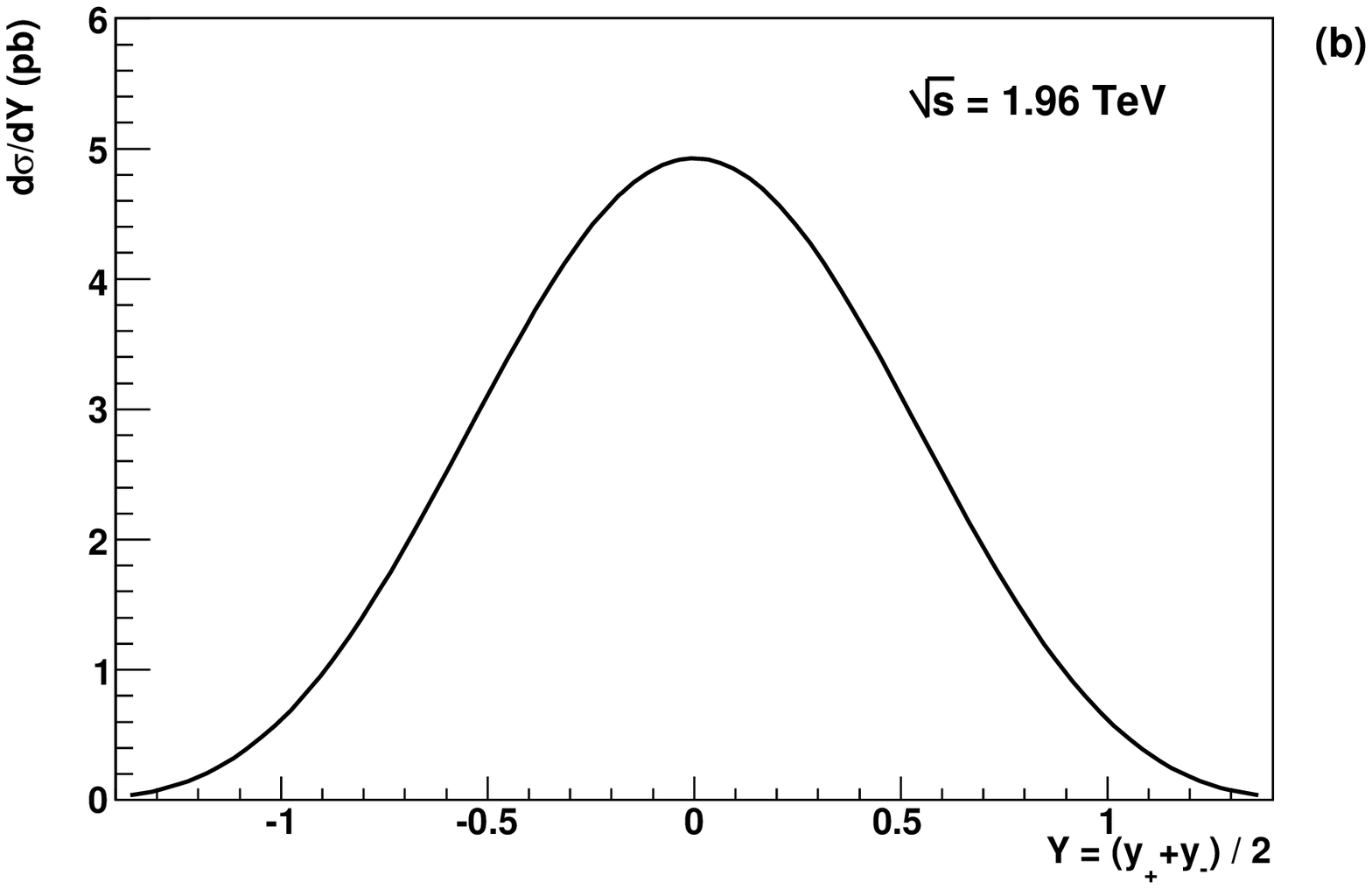}
\caption{\label{fig:average} (a) Differential QCD asymmetry in the 
average rapidity at Tevatron, $\sqrt{s}=1.96$~TeV. 
(b)~Average rapidity differential distribution.
Factorization and renormalization scales
are set to $\mu=m_t$.}
\end{center}
\end{figure}

The differential charge asymmetry of the single quark 
rapidity distribution is defined through 
\beq
A(y)=\frac{N_t(y)-N_{\bar{t}}(y)}{N_t(y)+N_{\bar{t}}(y)}~,
\label{eq:charge}
\eeq
where $y$ denotes the rapidity of the top (antitop) quark 
in the laboratory frame and $N(y)=d\sigma/dy$. 
Since $N_{\bar{t}}(y)=N_t(-y)$ as a consequence of charge 
conjugation symmetry, $A(y)$ can also be interpreted as a 
forward--backward asymmetry of the top quark. 
We have updated our previous analysis~\cite{mynlo}
by using the new value of the top quark mass,
$m_{t}=170.9 \pm 1.1~(\mathrm{stat}) \pm 
1.5~(\mathrm{sys})$~GeV~\cite{topmass},
and the new set of MSRT2004~\cite{Martin:2006qz} structure functions.
For the total charge asymmetry at $\sqrt{s}=1.96$~TeV we predict 
\beq
A =\frac{N_t(y\ge 0)-N_{\bar{t}}(y\ge 0)}
{N_t(y\ge 0)+N_{\bar{t}}(y\ge 0)} = 0.051(6)~,
\label{ourprediction}
\eeq
where different choices of the parton distribution 
functions, and different choices of the 
factorization and renormalization scales, $\mu=m_t/2$ and $\mu=2m_t$, 
and a variation of the top mass within the experimental error
have been considered. We have also included the 
contribution of the mixed QCD-electroweak interference that 
leads to an increase of the asymmetry as given by QCD by a 
factor $1.09$~\cite{mynlo}. The result is close to 
our earlier prediction: $A=4.8-5.8$\%~\cite{mynlo}.
Both the numerator and denominator are evaluated in LO. 
NLO corrections to the total $t\bar t$ production cross section 
are known to be large, around $30$\% or even larger~\cite{Bonciani:1998vc}. 
Applying a $K$-factor of $1.3$ to the denominator 
would reduce the asymmetry to $A=0.036(4)$.
In the absence of NLO corrections to the numerator 
we nevertheless stay with the LO approximation in both 
numerator and denominator, expecting the dominant 
corrections from collinear emission to cancel. 
From a more conservative point of view an uncertainty of 
around $30$\% has to be assigned to the prediction 
of the asymmetry. 

Let us now consider events where the rapidities 
$y_+$ and $y_-$ of both the top and antitop quarks
have been determined, and define a new differential distribution 
that leads to an enhancement of the charge asymmetry. We define 
\beq
Y=\frac{1}{2}(y_+ + y_-)~, 
\eeq
as average rapidity. Then, we consider the differential
pair asymmetry ${\cal A}(Y)$ for all events with fixed $Y$
as a function of $Y$: 
\beq
{\cal A}(Y) = \frac{\d N_\mathrm{ev.}(y_+>y_-) - N_\mathrm{ev.}(y_+<y_-)}
{\d N_\mathrm{ev.}(y_+>y_-) + N_\mathrm{ev.}(y_+<y_-)}~.
\eeq
In Figure~\ref{fig:average}(a) we show the differential pair 
asymmetry ${\cal A}(Y)$ versus the average rapidity for the Tevatron. 
The renormalization and factorization scales are set to 
$\mu=m_t$, and we use the LO MSRT2004~\cite{Martin:2006qz} 
parton distributions with three fixed flavours. 
Asymmetric contributions from flavour excitation processes, 
i.e. from $g q$($g \bar{q}$) collisions, are negligible. 
We obtain an almost flat positive asymmetry of the order of $7$\%.
For reference, we also plot in Figure~\ref{fig:average}(b)
the differential cross section as function of the 
average rapidity. For the integrated pair asymmetry we find 
\beq
{\cal A} = 
\frac{\d \int dY (N_\mathrm{ev.}(y_+>y_-) - N_\mathrm{ev.}(y_+<y_-))}
{\d \int dY (N_\mathrm{ev.}(y_+>y_-) + N_\mathrm{ev.}(y_+<y_-))}
=  0.078(9)~,
\label{eq:pair}
\eeq
where the error has been estimated as for the forward--backward 
asymmetry. Numerator and denominator are evaluated in leading order. 

The integrated pair asymmetry is equivalent to the definition 
of the asymmetry used in Refs.~\cite{Weinelt:2006mh,Hirschbuehl:2005bj}. 
The reason for the enhancement of the effect can be understood 
as follows: by defining the pair
asymmetry one essentially investigates the forward--backward 
asymmetry in the $t\bar{t}$ rest frame, where the forward--backward 
asymmetry amounts to $7-8.5$\%~\cite{mynlo}, depending on $\hat{s}$. 
This value is largely recovered by considering the pair 
asymmetry ${\cal A}(Y)$, independently of $Y$. In contrast events 
where both $t$ and $\bar{t}$ are produced with positive and 
negative rapidities do not contribute to the integrated 
forward--backward asymmetry $A$, which is therefore reduced 
to around $5$\%.

\section{Axigluon limits from Tevatron}

The interference between the gluon and axigluon induced amplitudes
respectively for the reaction $q\bar{q} \to t\bar{t}$ does not 
contribute to the production cross section. However, it generates a charge 
asymmetry that gives rise to a forward--backward asymmetry 
in $p\bar{p}$ collisions in the laboratory frame~\cite{Sehgal:1987wi}. 
The square of the axigluon amplitude is symmetric and contributes 
to the total cross-section, which will show a typical resonance 
peak in the top-antitop invariant 
mass distribution~\cite{Bagger:1987fz,Choudhury:2007ux}. 
While the interference term is suppressed by 
the squared axigluon mass $1/m_A^2$, the contribution of the 
square of the axigluon amplitude will be suppressed by $1/m_A^4$. 
It is therefore obvious that the forward--backward asymmetry is 
potentially sensitive to larger values of the axigluon mass
than the top-antitop dijet distribution . 
Gluon-gluon fusion is not affected by the axigluon exchange 
because there are no direct gluon-axigluon vertices with an odd 
number of axigluons~\cite{Bagger:1987fz} due to parity. 
The expressions that we use for the partonic Born 
cross-section are summarized in Appendix~\ref{ap:born}. 

\begin{figure}[th]
\begin{center}
\includegraphics[width=8cm]{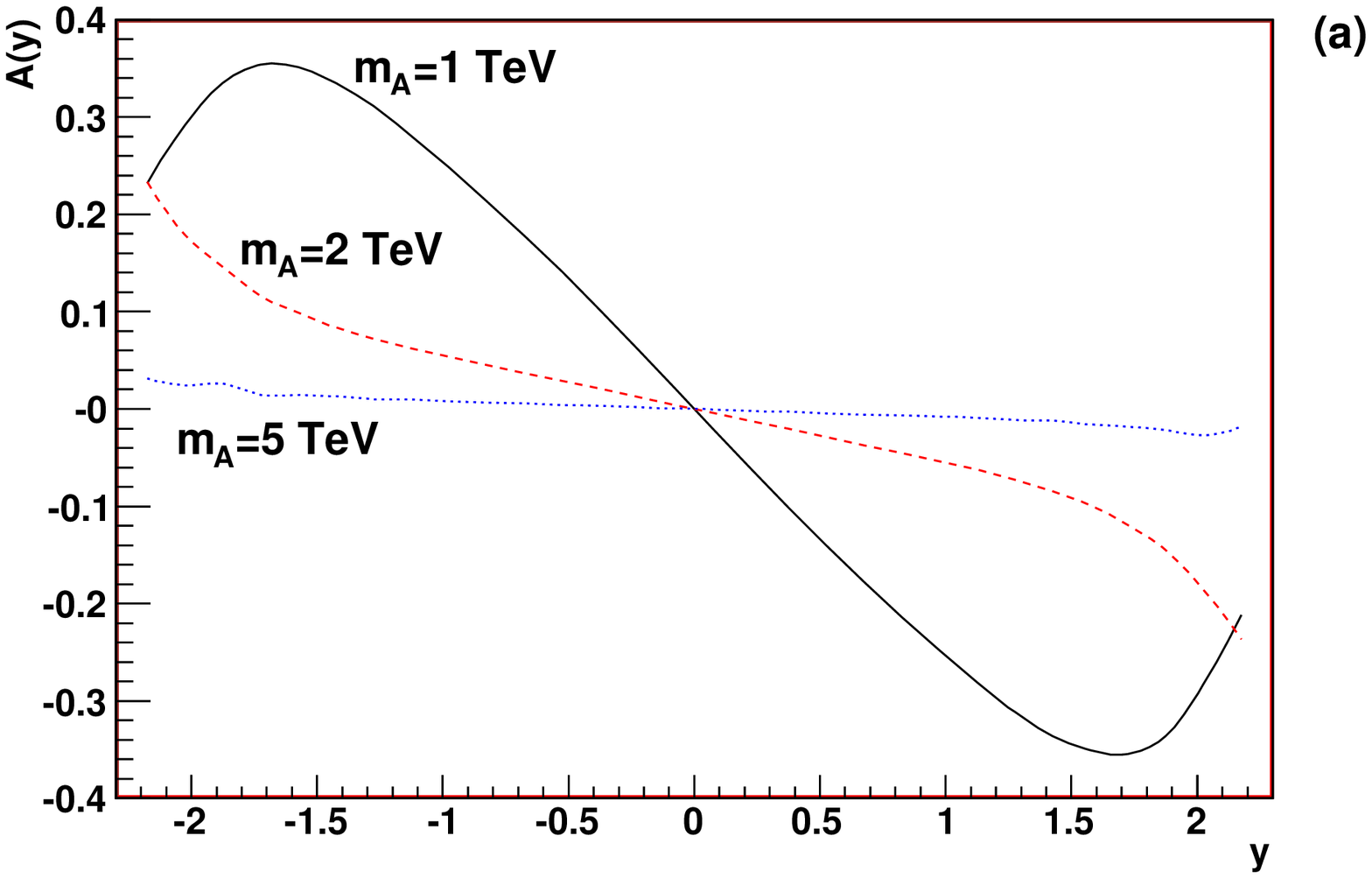}
\includegraphics[width=8cm]{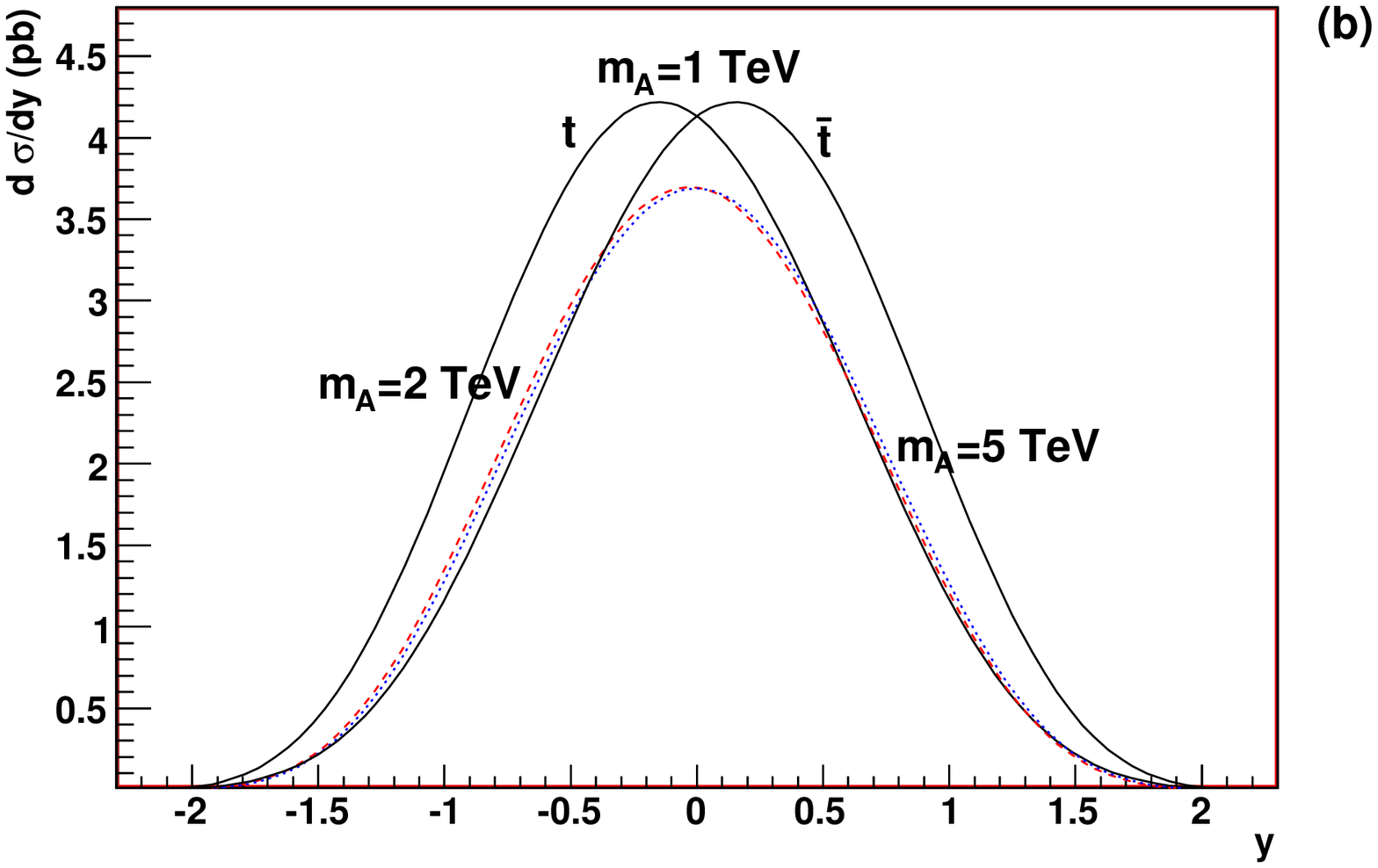}
\caption{\label{fig:tevatron} (a) Differential top quark charge 
asymmetry in $p\bar p$ collisions at Tevatron, $\sqrt{s}=1.96$~TeV, 
for different values of the axigluon mass. 
(b) Rapidity distribution of top and antitop quarks for $m_A=1$~TeV. 
For $m_A=2$~TeV (dashed) and $5$~TeV (dotted) 
we only represent the top quark distribution.
Factorization and renormalization scales
are set to $\mu=m_t$.}
\end{center}
\end{figure}

\begin{figure}[th]
\begin{center}
\includegraphics[width=8cm]{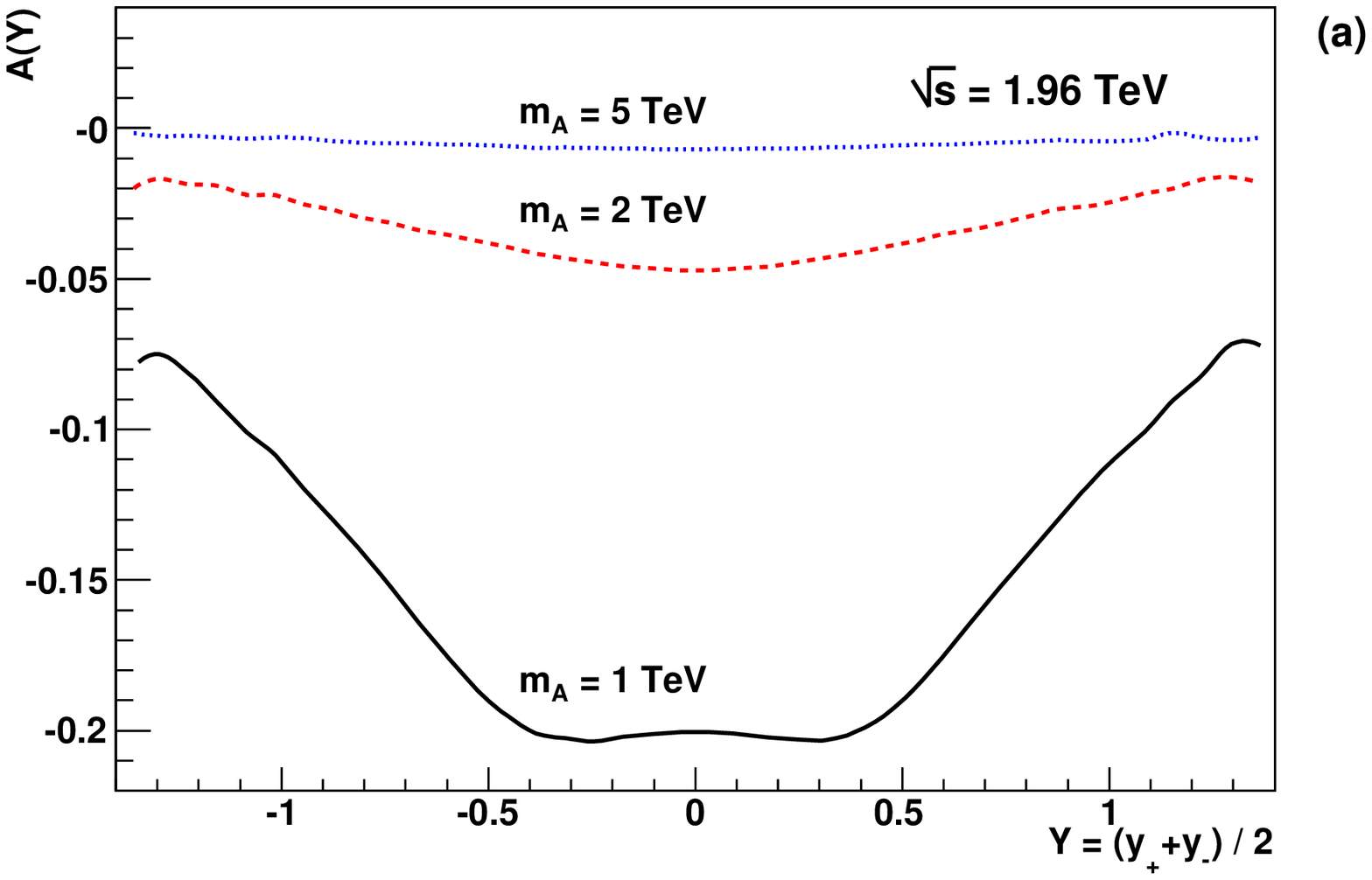}
\includegraphics[width=8cm]{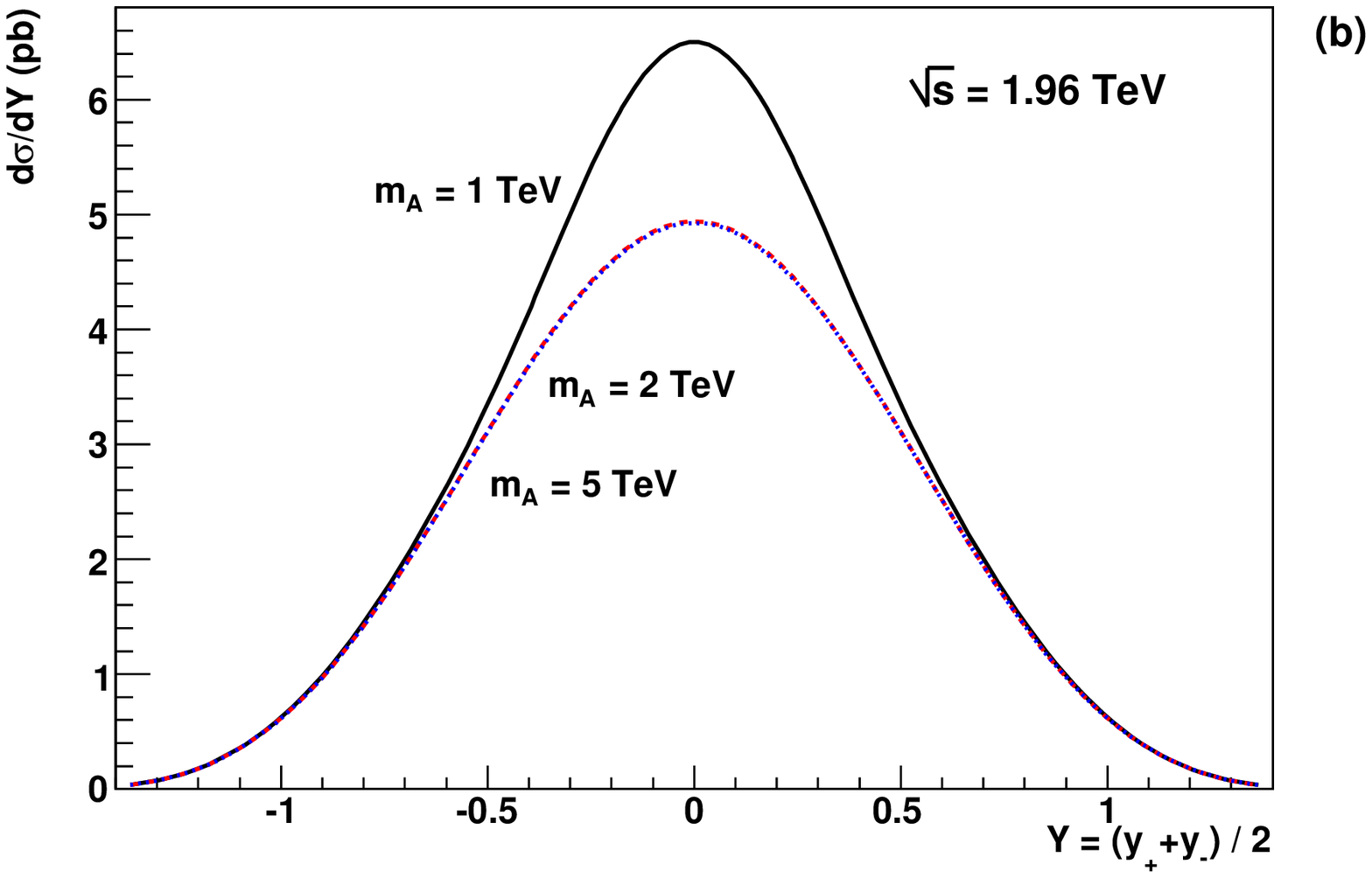}
\caption{\label{fig:Ytevatron} (a) Differential pair 
asymmetry in $p\bar p$ collisions at Tevatron, $\sqrt{s}=1.96$~TeV, 
for different values of the axigluon mass. 
(b) Corresponding differential cross section $d\sigma/dY$.
Factorization and renormalization scales are set to $\mu=m_t$.}
\end{center}
\end{figure}

\begin{table}[th]
\begin{center}
\begin{tabular}{ccccc} \hline
& QCD & $m_A=1$~TeV & $m_A=2$~TeV & $m_A=5$~TeV \\ \hline
$A_{\rm{FB}}=A$ 
& $0.051(6)$ & $-0.133(9)$  & $-0.027(2)$ & $-0.0041(3)$ \\ 
${\cal A}$                
& $0.078(9)$ & $-0.181(11)$ & $-0.038(3)$ & $-0.0058(4)$ \\ \hline 
\end{tabular}
\caption{Forward--backward and pair asymmetries
at Tevatron, $\sqrt{s}=1.96$~TeV, for different values of 
the axigluon mass.}
\label{tab:axiasymmetries}
\end{center}
\end{table}

\begin{figure}[th]
\begin{center}
\includegraphics[width=8cm]{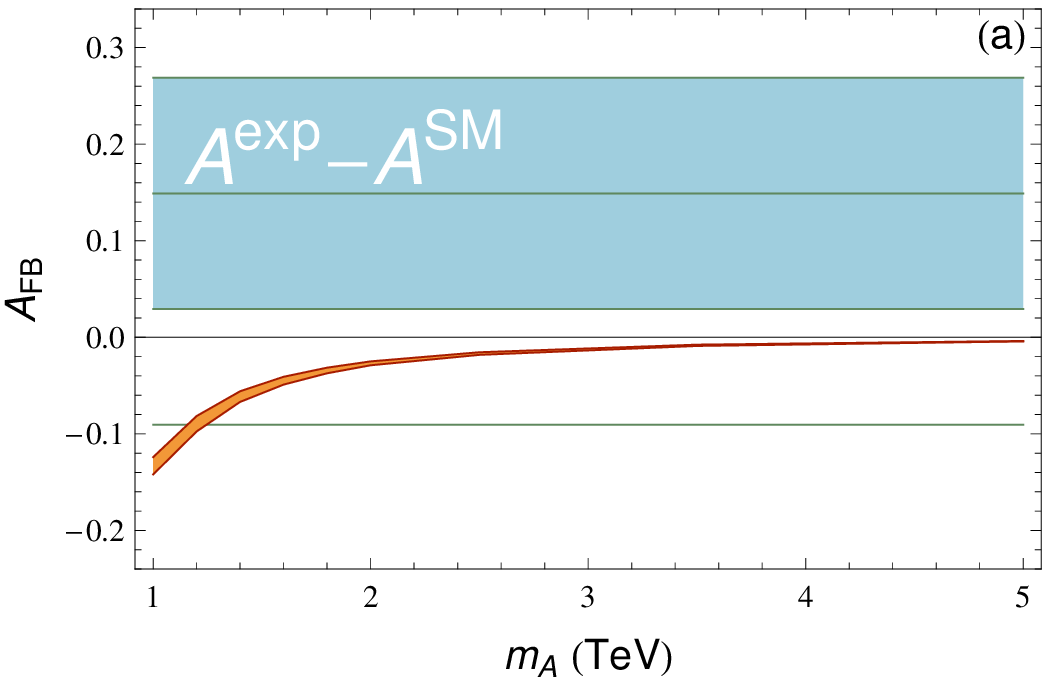} \hspace{.5cm}
\includegraphics[width=8cm]{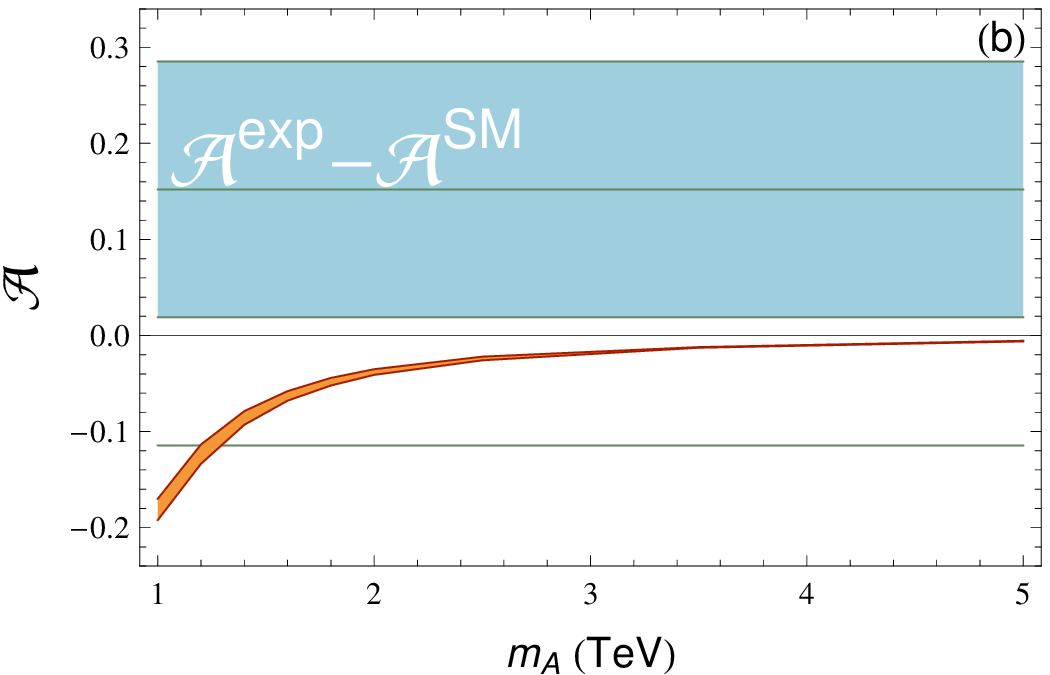} 
\caption{\label{fig:axiasym} 
Comparison of the top quark forward--backward (a) and pair (b) 
asymmetry generated by the exchange of a massive axigluon to the 
experimental measurement at Tevatron once the theoretical 
prediction in the SM is subtracted. The $2\sigma$ contour is also 
showed.}
\end{center}
\end{figure}

The interference between the amplitudes for production of $t\bar{t}$
through a gluon and an axigluon, respectively, vanishes upon 
integration over any charge symmetric part of the phase space 
and thus does not contribute to the production cross-section.
However, the charge asymmetry resulting from this interference may 
well be more sensitive to the existence of axigluons than the 
contribution of the squared amplitude, in particular in the 
case of large $m_A$, and will, furthermore, be a characteristic 
consequence of its axial nature. 

In Fig.~\ref{fig:tevatron}(a) the differential charge 
asymmetry $A(y)$ at the Tevatron is shown 
for different values of the axigluon mass
as a function of the top quark rapidity. 
The QCD induced piece is not included in this figure. 
In Fig.~\ref{fig:tevatron}(b) the top and antitop quark rapidity
distributions are shown, again at LO,
for $m_A=1$~TeV (for larger values of the 
axigluon mass we only plot the top quark distribution). 
Due to the axigluon contribution  
the top quark distribution is shifted to negative values 
of the rapidity while the antitop quark distribution prefers
positive values. The corresponding curves for the pair 
asymmetry ${\cal A}(Y)$ are shown in Fig.~\ref{fig:Ytevatron}(a), 
the differential pair cross section $d\sigma/dY$ 
in Fig.~\ref{fig:Ytevatron}(b).
In Table~\ref{tab:axiasymmetries} we present 
the prediction for the integrated forward--backward asymmetry
for different values of the axigluon mass.
We also give values for the integrated pair asymmetry
${\cal A}$ as defined in \Eq{eq:pair}. The uncertainty of the 
predictions is estimated as before. The size of the pair asymmetry 
is larger than the forward--backward asymmetry. 
The enhancement factor is however somewhat smaller than in QCD. 
 
As explained in the introduction, the forward--backward asymmetry 
$A_{\rm{FB}}$ and the integrated pair asymmetry have already been measured at 
the Tevatron~\cite{Weinelt:2006mh,Hirschbuehl:2005bj,Schwarz:2006ud}.
The uncertainty of these preliminary results
is still very large, and statistically dominated. The different 
experimental analyses are based on the same data sample and 
give a very similar result for the inclusive asymmetry. 
The subsequent discussion is thus of qualitative nature only. 
In Fig.~\ref{fig:axiasym}(a) and~\ref{fig:axiasym}(b) 
we compare the experimental 
measurements, with the SM contribution subtracted 
(\Eq{ourprediction} and \Eq{eq:pair} respectively), 
to the asymmetry generated by the axigluon. At the two sigma 
level axigluon masses below $m_A=1.2$~GeV are excluded. 
Both asymmetries give a very similar lower bound. 
It is clear that a more accurate experimental measurement 
and theoretical prediction will allow to further constrain  
the axigluon mass. However, at present the largest uncertainty 
by far is of experimental origin, and it will be reduced with 
more statistics. In this case, we will be in the interesting 
situation where axigluon masses that are not accessible at  
Tevatron through the study of the dijet cross-section can be 
excluded by the forward--backward or the pair asymmetries.

\section{QCD and axigluon induced asymmetries at the LHC}

\begin{figure}[ht]
\begin{center}
\includegraphics[width=8cm]{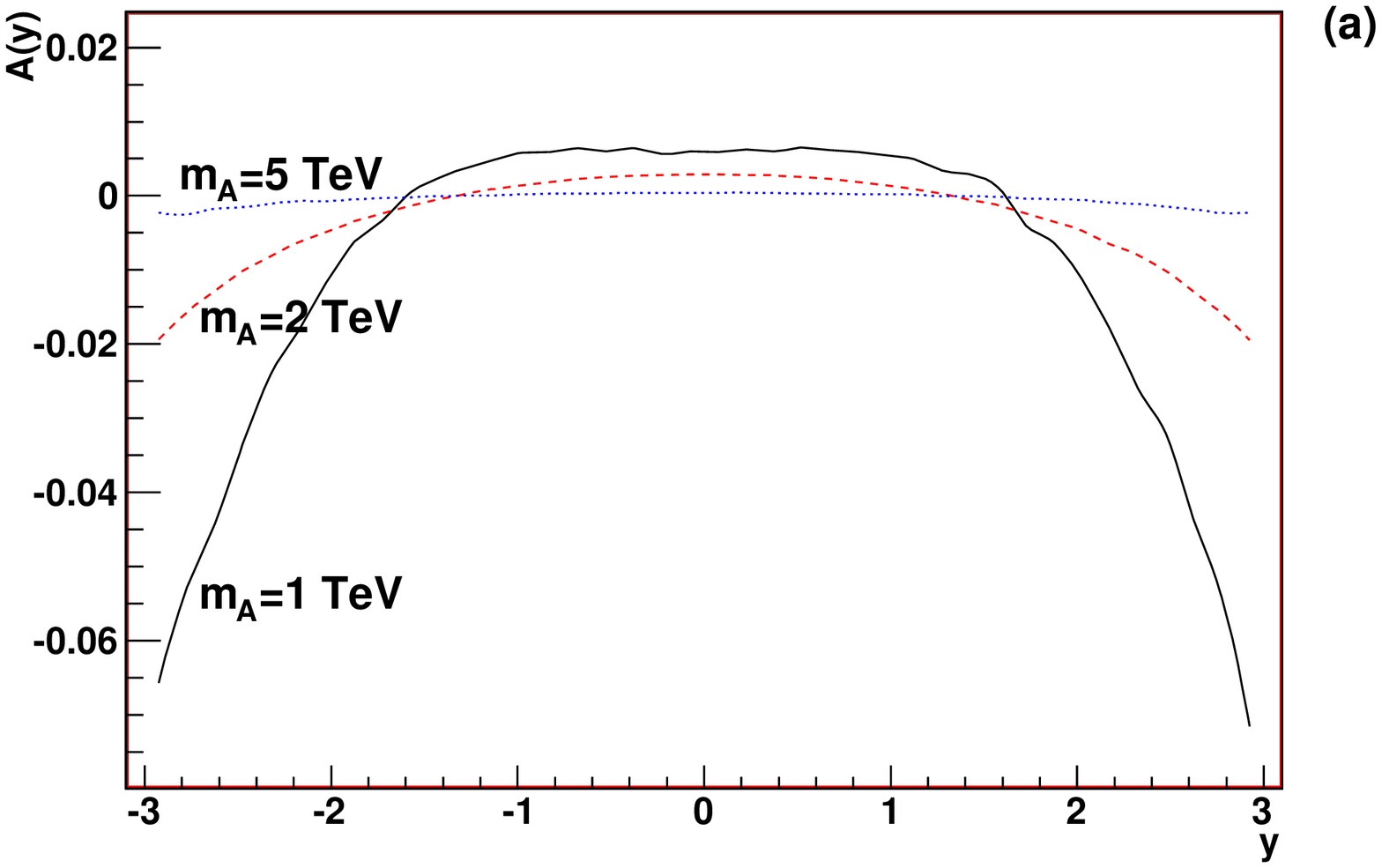}
\includegraphics[width=8cm]{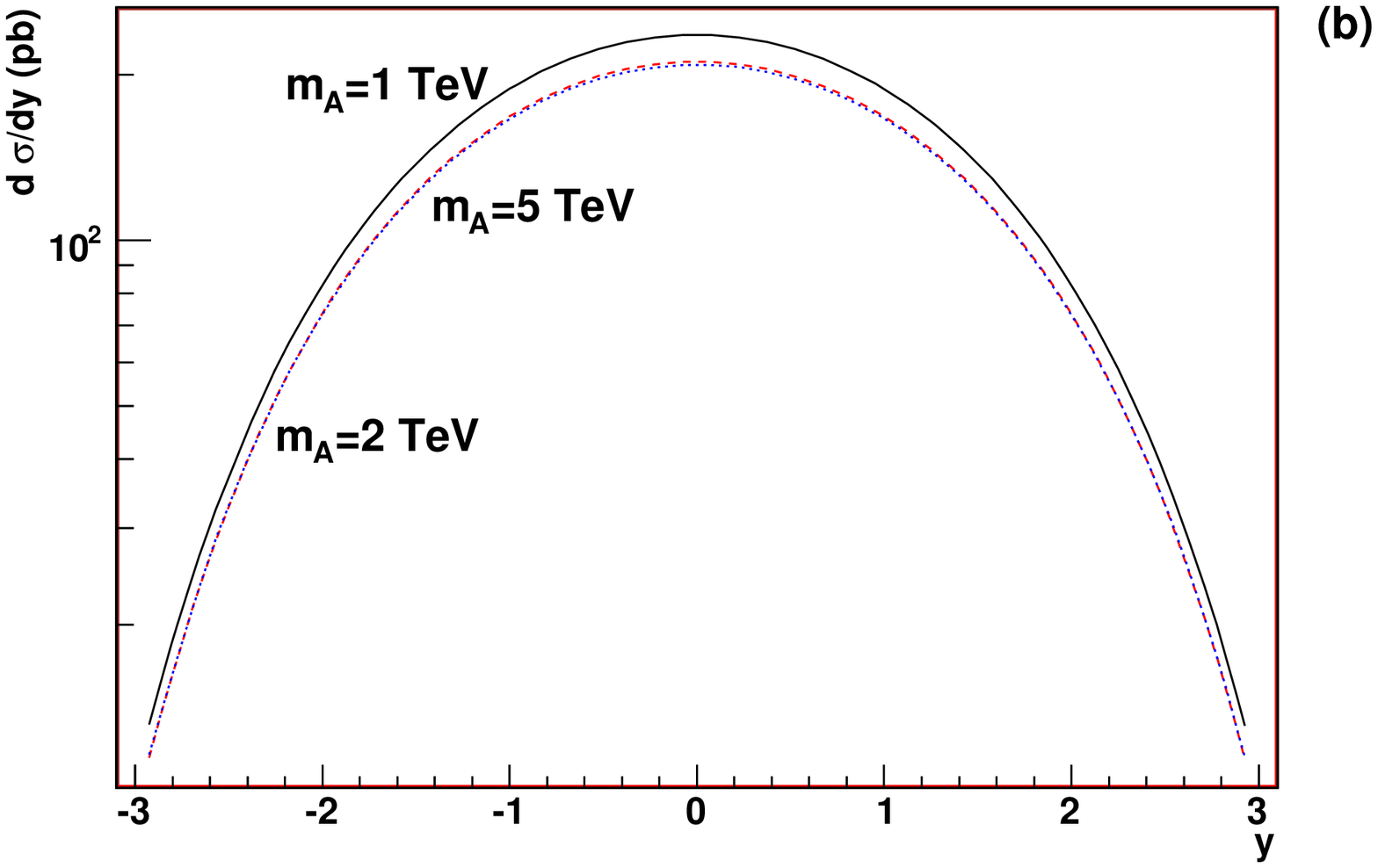}
\caption{\label{fig:lhc} (a) Differential top quark charge 
asymmetry in $pp$ collisions at LHC, $\sqrt{s}=14$~TeV, 
for different values of the axigluon mass. 
(b) Rapidity distribution of top quarks  
for different values of the axigluon mass.
Factorization and renormalization scales are set to $\mu=m_t$.}
\end{center}
\end{figure}

Top quark production at LHC is forward--backward symmetric in the 
laboratory frame as a consequence of the symmetric colliding 
proton-proton initial state. Furthermore, the total cross section 
is dominated by gluon-gluon fusion and thus the charge asymmetry 
generated from the $q\bar{q}$ and $gq$ ($g\bar{q}$) reactions is
negligible in most of the kinematic phase-space. The effect 
can be studied nevertheless by selecting appropriately chosen 
kinematic regions. At LHC the QCD asymmetry predicts a slight 
preference for centrally produced antitop quarks, with top quarks 
more abundant at very large positive and negative rapidities~\cite{mynlo}. 
The charge asymmetry as defined in \Eq{eq:charge} is, however, 
only sizable in regions with low event rates and large rapidities, 
where the experimental observation might be difficult. 
In Fig.~\ref{fig:lhc}(a) we present the differential charge 
asymmetry generated by the exchange of an axigluon 
as a function of rapidity, choosing $m_A=1,2$ and $5$~TeV. 
We observe a similar behaviour as predicted by 
QCD but with the opposite sign; top quarks are slightly
more abundant in the central region, while a sizable negative 
asymmetry is found at large values of the rapidity. 
For $m_A=5$~TeV the asymmetry is almost negligible 
throughout, a consequence of the relatively small 
$t\bar{t}$ mass and the correspondingly strong suppression 
of the axigluon amplitude.

\begin{figure}[th]
\begin{center}
\includegraphics[width=8cm]{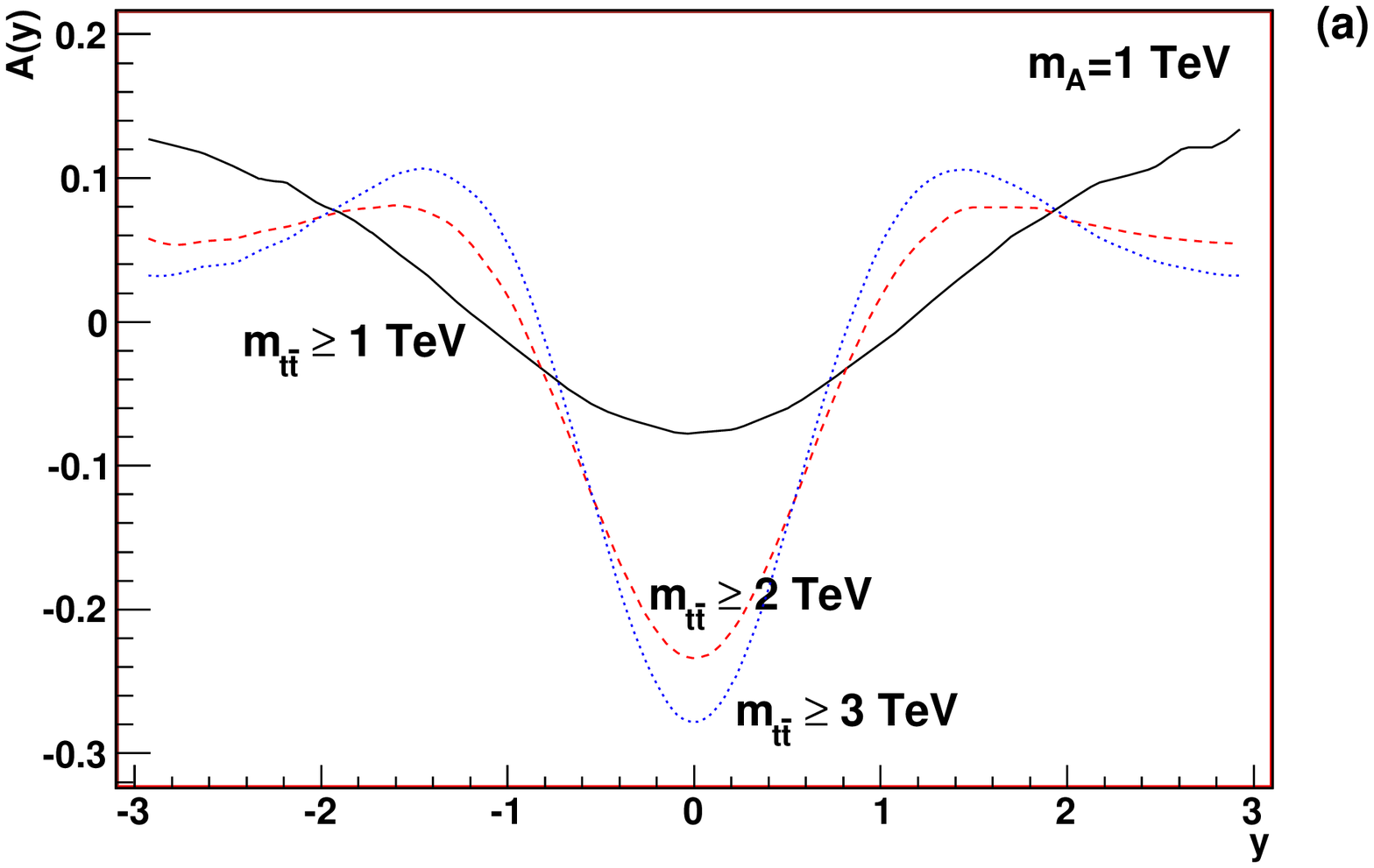}
\includegraphics[width=8cm]{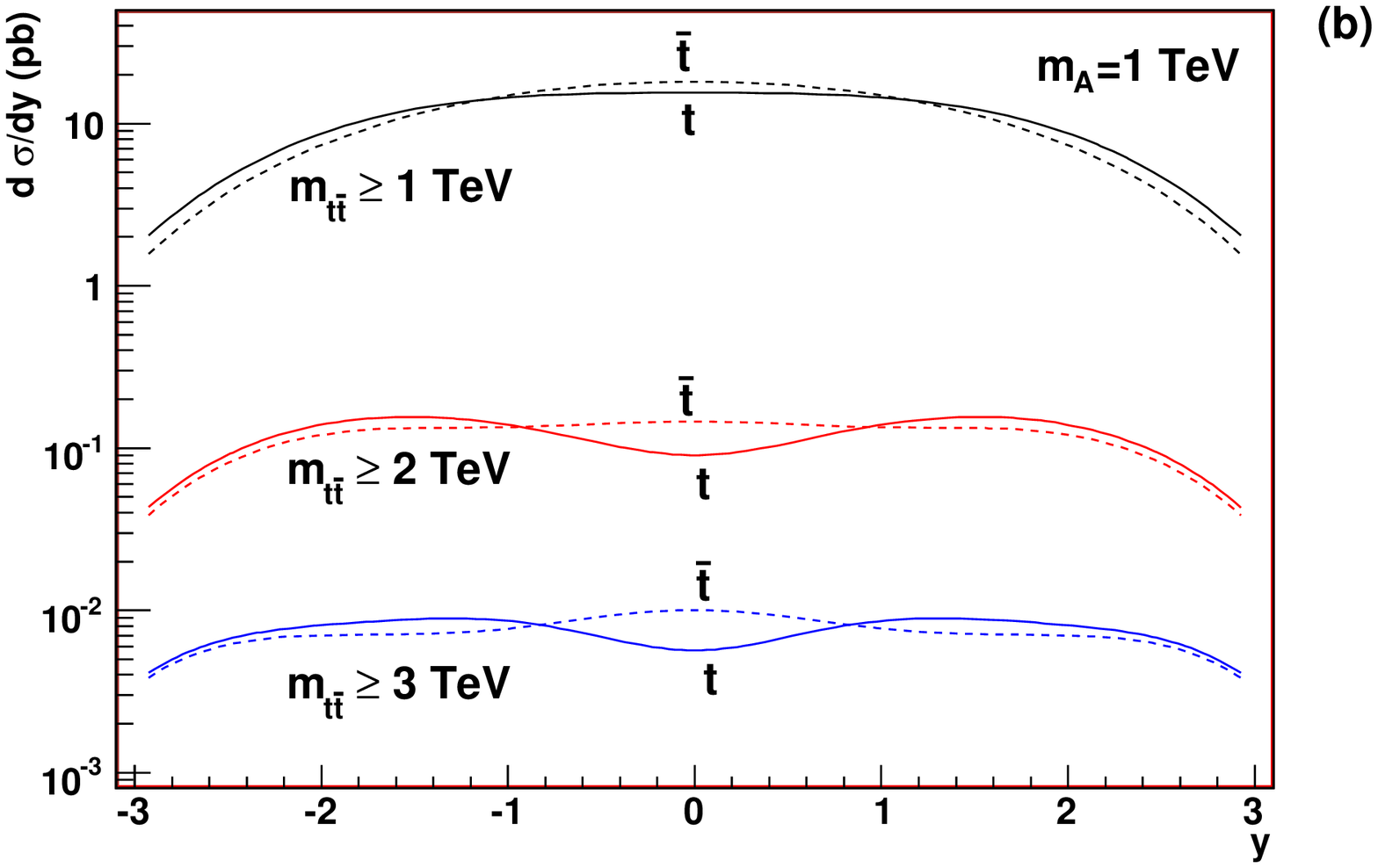}
\caption{\label{fig:lhc1Tev} (a) Differential top quark charge 
asymmetry at LHC generated by the exchange of a massive axigluon with 
$m_A=1$~TeV for samples with invariant top-antitop quark masses
larger than $1$, $2$ and $3$~TeV respectively. 
(b) Corresponding rapidity distributions of top quarks (solid) and 
antitop quarks (dashed). Factorization and renormalization scales
are set to $\mu=m_t$.}
\end{center}
\end{figure}

\begin{figure}[th]
\begin{center}
\includegraphics[width=8cm]{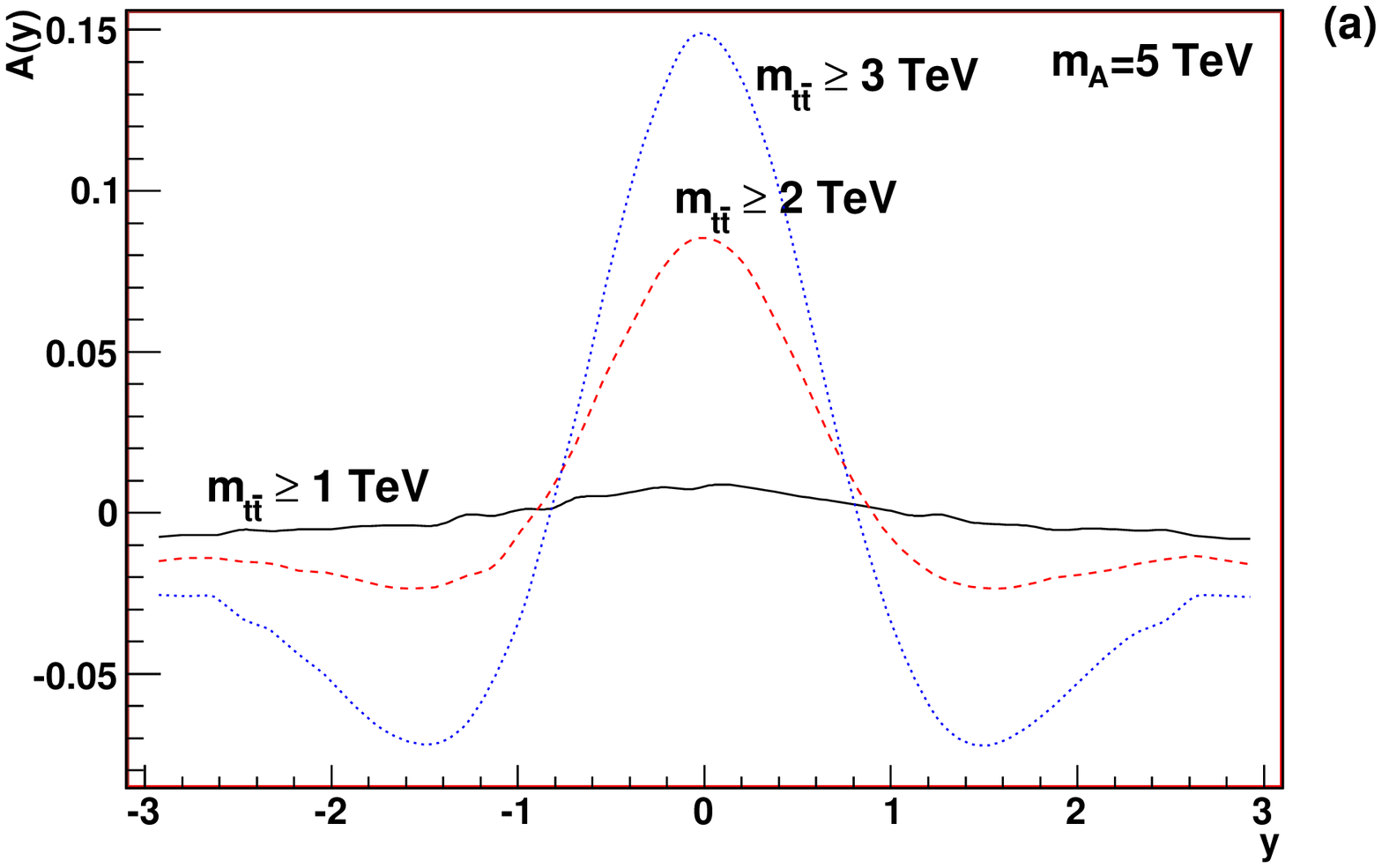}
\includegraphics[width=8cm]{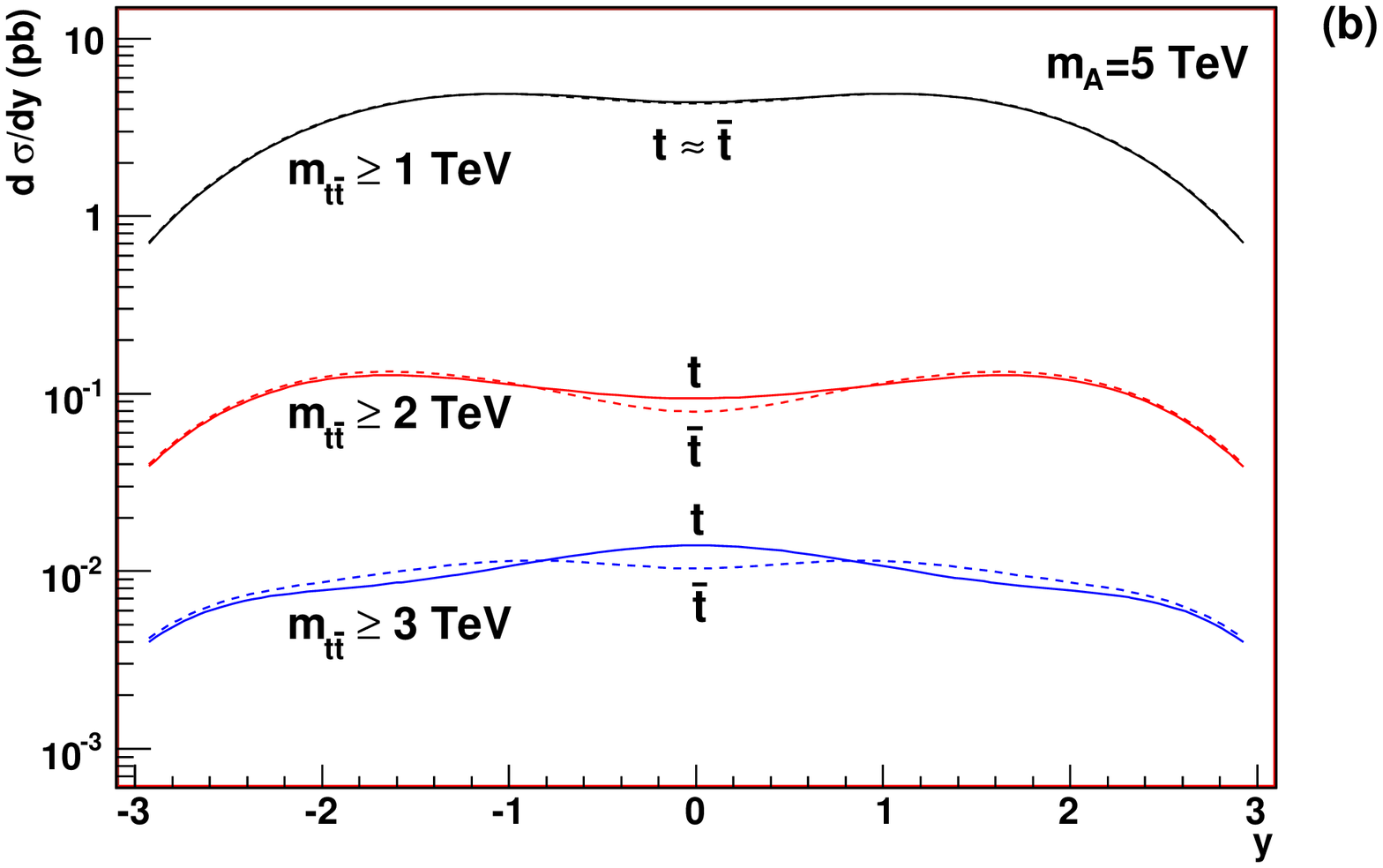}
\caption{\label{fig:lhc5Tev} (a) Differential top quark charge 
asymmetry at LHC generated by the exchange of a massive axigluon with 
$m_A=5$~TeV for samples with invariant top-antitop quark masses
larger than $1$, $2$ and $3$~TeV respectively. 
(b) Corresponding rapidity distributions of top quarks (solid) and 
antitop quarks (dashed). Factorization and renormalization scales
are set to $\mu=m_t$.}
\end{center}
\end{figure}

We shall therefore analyze the effect of selecting samples with 
high invariant masses of the top-antitop quark pair. Those samples 
should have a higher amount of $q\bar{q}$ induced events, 
and furthermore, an enhanced axigluon amplitude even for large $m_A$.  
Thus a sizable asymmetry is expected, although at the 
price of reducing the total event rate. This should not 
be a problem at LHC due to the huge top-antitop quark 
yields. 
In Fig.~\ref{fig:lhc1Tev}(a) we show the differential 
charge asymmetry generated by axigluons of mass $m_A=1$~TeV
for samples with invariant masses larger than $1$, $2$ and 
$3$~TeV respectively. A large asymmetry with a maximum in the 
central region ranging from $-8$\% to $-30$\%  is predicted. 
In Fig.~\ref{fig:lhc1Tev}(b) we show the corresponding top and 
antitop quark rapidity distributions. The total single inclusive
cross sections are equal but the rapidity distributions are 
not, with antitop quarks more abundant in the central 
region. 
Similar plots are presented in Fig.~\ref{fig:lhc5Tev}
for an axigluon mass of $m_A=5$~TeV, where now the generated 
asymmetry becomes positive due to the larger axigluon mass. 
In the former case the process is dominated by contributions 
with $\hat s > m_A^2$, in the latter case with $\hat s < m_A^2$, 
which implies a change in the sign of the axigluon propagator.

\begin{figure}[ht]
\begin{center}
\includegraphics[width=9cm]{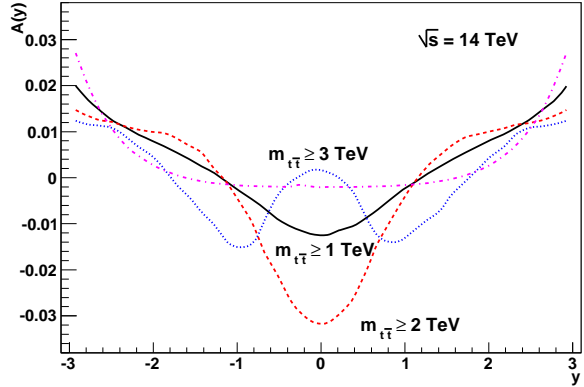}
\centering
\caption{\label{fig:lhcQCDasym} Differential QCD charge 
asymmetry at LHC for top-antitop quark invariant masses larger than 
$2m_t$ (dotted-dashed),
$1$~TeV (solid), $2$~TeV (dashed), and $3$~TeV (dotted) respectively. 
Factorization and renormalization scales are set to $\mu=m_t$.}
\end{center}
\end{figure}

In order to quantify the difference in the rapidity distribution 
globally we define a new charge asymmetry where only the the 
central region is taken into account:
\beq
A_C(y_C) = \frac{\sigma_t(|y|\le y_C)-\sigma_{\bar{t}}(|y|\le y_C)}
{\sigma_t(|y|\le y_C)+\sigma_{\bar{t}}(|y|\le y_C)}~.
\eeq
Notice that $A_C(y_C)$ vanishes if the whole rapidity spectrum
is integrated. Looking into Figs.~\ref{fig:lhc1Tev}(a)
and~\ref{fig:lhc5Tev}(a) we expect that 
the maximum is reached at about $y_C=1$. 
We thus report in Table~\ref{tab:axiasymmetrieslhc} the 
values of the central asymmetry, integrated up to $y_C=1$
for different values of the axigluon mass and for different 
values of the cut in the top-antitop invariant mass.
The single inclusive top quark cross section in the central 
region is also listed (the corresponding antitop quark cross 
section can be deduced from the asymmetry). 
We also give in Table~\ref{tab:axiasymmetrieslhc} the value of
the QCD asymmetry in the central region, and in 
Fig.~\ref{fig:lhcQCDasym} the corresponding
differential charge asymmetry. The central QCD charge 
asymmetry is never larger than $2$\%, and the reduction 
of the asymmetry for invariant masses above $2$~TeV is due 
to a larger contribution of the flavour excitation processes 
that partly compensates the asymmetry generated by the $q\bar{q}$
events.

With $10$~fb$^{-1}$ integrated luminosity about 
2000-2600 top quark events with $m_{t\bar{t}}\ge 2$~TeV are 
expected to be produced at LHC in the central region ($y_C=1$), 
and a central charge asymmetry of about $-2$\% is predicted 
from QCD alone. For $m_A= 5$~TeV the axigluon-induced asymmetry 
amounts to $+3\%$.
For the same axigluon mass the total cross-section for 
$m_{t\bar{t}}\ge 4.5$~TeV is only $0.013$~pb, one order of magnitude 
smaller than the central cross section for $m_{t\bar{t}}\ge 2$~TeV 
and $y_C=1$. It is thus clear that even at LHC where no forward--backward 
asymmetry is generated, the central charge asymmetry can be used to 
probe larger axigluon masses than the top-antitop invariant mass 
spectrum~\cite{Bagger:1987fz,Choudhury:2007ux}.

\begin{table}[th]
\begin{center}
\begin{tabular}{cccccc} \hline
& & QCD & $m_A=1$~TeV & $m_A=2$~TeV & $m_A=5$~TeV \\ \hline
$\sqrt{\hat{s}} \ge 1$~TeV 
& $A_C(y_C=1)$ 
& $-0.0086 (4)$ & $-0.055 (4)$ & $0.025 (3)$ & $0.002(1)$\\
& $\sigma_t(|y|\le 1)$  
& $9.7 (2.7)$~pb& $34(4)$~pb & $15(2)$~pb & $11 (2)$~pb \\ \hline 
$\sqrt{\hat{s}} \ge 2$~TeV 
& $A_C(y_C=1)$ 
& $-0.0207(14)$ & $-0.10 (2)$ & $-0.048 (5)$ & $0.031(9)$  \\
& $\sigma_t(|y|\le 1)$  
& $0.19 (6)$~pb& $0.28 (8)$ pb & $1.7 (2)$~pb & $0.26 (7)$ pb \\ \hline 
$\sqrt{\hat{s}} \ge 3$~TeV 
& $A_C(y_C=1)$ 
& $-0.0151 (7)$ & $-0.10 (3)$ & $-0.11 (2)$ & $0.057 (13)$\\
& $\sigma_t(|y|\le 1)$  
& $0.011 (4)$~pb & $0.019 (6)$~pb & $0.024 (7)$~pb & $0.031 (8)$~pb\\ \hline 
\end{tabular}
\caption{Central charge asymmetry at LHC, $\sqrt{s}=14$~TeV, 
for different values of the axigluon mass and different cuts 
in the invariant mass of the top-antitop pair. We also present 
the LO prediction for the top quark cross section in the 
central region.}
\label{tab:axiasymmetrieslhc}
\end{center}
\end{table}

\section{Summary}

We have updated our previous analysis of the forward--backward and 
the charge asymmetry in top quark production at hadron colliders. 
We have also proposed a new observable , the pair asymmetry, 
where the effect at the Tevatron is enhanced by about a factor~$1.5$.
Top quark production at the LHC is obviously forward--backward 
symmetric. Restricting the sample to events with large 
invariant $t\bar{t}$ mass and to $t$ and $\bar{t}$ final states 
with rapidity below one a difference of $1-2$\% in $t$ versus 
$\bar{t}$ production cross section is observed. 

The analysis has been extended to the asymmetry induced by 
axigluon contributions to the amplitude, which is the most 
important manifestation of such an exotic object. Already 
present preliminary results on the charge asymmetry from the
Tevatron lead to an limit on the axigluon mass of around~$1.2$~TeV.
The analysis has been extended to the LHC. Restricting the event 
sample to regions of large $t\bar{t}$ mass with $q\bar{q}$ 
induced production, large axigluon masses can be explored.

\section*{Acknowledgements}

We thank Jeannine Wagner, Julia Weinelt
and Dominic Hirschb\"uhl for very useful discussions.  
One of the authors (JK) is grateful to the Max-Planck-Institut f\"ur
Physik, where part of this paper was completed, for its hospitality.
Work partially supported by 
Consejo Superior de Investigaciones Cient\'{\i}ficas (CSIC) under
grant PIE 200650I247, Ministerio de Educaci\'on y Ciencia (MEC) under
grant FPA2004-00996, Generalitat Valenciana under grant GVACOMP2007-156, 
BMBF grant No. 05HT4VKA/3, and European Commission MRTN FLAVIAnet 
under contract MRTN-CT-2006-035482.



\appendix

\section{Born cross-section}
\label{ap:born}

The Born cross-section for $q\bar{q}$ fusion in the presence of 
an axigluon vector resonance reads
\bea
\frac{d\sigma^{q\bar{q}\rightarrow t \bar{t}}}{d\cos \hat{\theta}} = 
\alpha_s^2 \: \frac{T_F C_F}{N_C} \:  
\frac{\pi \beta}{2 \hat{s}}
\left( 1+c^2+4m^2 +
\frac{4 \, c \, \hat{s} (\hat{s}-m_A^2) + \hat{s}^2 \, (\beta^2+c^2)}
{(\hat{s}-m_A^2)^2+m_A^2 \Gamma_A^2}
\right)~,
\label{eq:bornqq}
\eea
where $\hat{\theta}$ is the polar angle of the top quark with respect 
to the incoming quark in the center of mass rest frame, 
$\hat{s}$ is the squared partonic invariant mass,
$T_F=1/2$, $N_C=3$ and $C_F=4/3$ are the color factors,
$\beta = \sqrt{1-4m^2}$ is the velocity of the top quark, 
with $m=m_t/\sqrt{\hat{s}}$, and $c = \beta \cos \hat{\theta}$.
Our result in \Eq{eq:bornqq} agrees with the expression of 
Ref.~\cite{Choudhury:2007ux}, and therefore confirms the 
disagreement with respect to Eq.(2) of Ref.~\cite{Sehgal:1987wi}.
The term odd in $c$ in \Eq{eq:bornqq} is due to the interference
between the gluon and the axigluon amplitudes. 
The decay width of the axigluon is given 
by~\cite{Bagger:1987fz,Choudhury:2007ux}:
\beq
\Gamma_A \equiv \sum_q \Gamma (A \to q\overline{q}) 
\approx \frac{\alpha_{s}\, m_A \, T_F}{3} 
\left[5+\left(1-\frac{4m_t^{2}}{m_A^{2}}\right)^{3/2}\right]~.
\eeq
Because of parity there are no gluon-axigluon vertices with an odd 
number of axigluons~\cite{Bagger:1987fz}, and therefore the Born 
gluon-gluon fusion cross-section is the same as in the SM~\cite{mynlo}:
\beq
\frac{d\sigma^{gg\rightarrow t \bar{t}}}{d\cos \hat{\theta}} = 
\alpha_s^2 \: \frac{\pi \beta}{2 \hat{s}}  
\left(\frac{1}{N_C(1-c^2)}-\frac{T_F}{2C_F}\right)
\left(1 + c^2 +8 m^2-\frac{32 m^4}{1-c^2}\right)~.
\eeq


\end{document}